\documentclass[useAMS,usenatbib,usegraphicx]{mn2e}
\usepackage[english]{babel}
\usepackage{times}
\usepackage[applemac]{inputenc}
\usepackage{amsmath}
\usepackage{fancybox}
\usepackage{fancyhdr}
\usepackage{color}
\usepackage{amssymb} 
\usepackage{inputenc} 
\usepackage{enumerate}
\usepackage{graphicx}
\usepackage{natbib}
\usepackage{lscape}

\newcommand{\apj}{ApJ}

\newcommand{\apjl}{ApJL}
\newcommand{\mnras}{MNRAS}
\newcommand{\aap}{A\&A}

\def\4u{4U~0614+09}
\begin{document}

\title{kHz Quasi-Periodic Oscillations in the low-mass X-ray binary 4U 0614+09}
\author[Martin Boutelier, Didier Barret and  M. Coleman Miller]{Martin Boutelier,$^{1,2}$\thanks{Email : martin.boutelier@cesr.fr}  Didier Barret$^{1,2}$ and  M. Coleman Miller$^3$ \\
$^1$Universit\'e de Toulouse (UPS) \\
$^2$Centre National de la Recherche Scientifique, Centre d'Etude Spatiale des Rayonnements, UMR 5187, 9 av. du Colonel Roche, BP 44346, 31028 Toulouse Cedex 4 \\
$^3$Department of Astronomy and Maryland Astronomy Center for Theory
and Computation, University of Maryland, College Park, MD 20742-2421, USA}
\date{}
\maketitle
\begin{abstract} 
We report on a comprehensive analysis of the kilohertz ($\geq 300$ Hz)
quasi-periodic oscillations (kHz QPOs) detected from the neutron star
low-mass X-ray binary \4u with the {Rossi X-ray Timing Explorer}
(RXTE). With a much larger data
set than previously analyzed (all archival data from February 1996 
up to October 2007), we first
investigate the reality of the 1330 Hz QPO reported by
\citet{van-Straaten:2000zl}. This QPO would be of particular interest since it
has the highest frequency reported for any source. A thorough analysis of the
same observation fails to confirm the detection. On the other hand,
over our extended data set, the highest QPO frequency we measure for
the upper kHz QPO is at $\sim 1224$ Hz; a value which is fully
consistent with the maximum values observed in similar systems.
Second, we demonstrate that the frequency dependence of the quality
factor ($Q=\nu/\Delta\nu$) and amplitude of the lower and upper kHz
QPOs follow the systematic trends seen in similar
systems \citep{Barret:2006it}. In
particular, \4u shows a drop of the quality factor of the lower kHz QPO above
$\sim 700$ Hz. If this is due to an approach to the innermost stable
circular orbit, it implies a neutron star mass of $\rm \sim
1.9~M_\odot$. Finally, when analyzing the data over fixed durations, we have found a gap in the frequency distribution of the upper QPO, associated with a local minimum of its amplitude. A similar gap is not present in the distribution of the lower QPO frequencies, suggesting some cautions when interpreting frequency ratio distributions, based on the occurrence of the lower QPO only.
\end{abstract}

\section{Introduction}
\label{intro}

Kilohertz quasi-periodic brightness oscillations (kHz QPOs) have been reported from \4u by \cite{Ford:1997hl,van-Straaten:2000zl,van-Straaten:2002db,Barret:2006it,Mendez:2006qq}. The peculiar
properties of its kHz QPOs motivate the present work for three main
reasons. First, among kHz QPO sources, \4u holds the record for the
highest claimed QPO frequency, 1330~Hz \citep{van-Straaten:2000zl}, 
whereas in most sources the maximum
frequency for the upper kHz QPO lies around 1200 Hz. This
is of particular importance because it sets the most stringent
constraints on the mass and radius of the NS, under the assumption that
1330~Hz is an orbital frequency \citep{Miller:1998ek}.  Unfortunately,
\4u tends to have a low count rate and broad QPOs compared to similar
sources, hence its QPOs are challenging to characterize.

Second, \citet{Barret:2006it} have performed a systematic study
of the quality factor of the lower and upper kHz QPOs in six
systems: 4U~1636-536, 4U~1608-522, 4U~1735-44, 4U~1728-34,
4U~1820-303 and \4u. Using data available in the RXTE archive at
the end of 2004, they found that all the sources except \4u
showed evidence of a drop in the quality factor of
their lower kHz QPOs at high frequency. For \4u only the rising
part of the quality factor versus frequency curve was reported.
The sudden drop is consistent with what is expected if it is
produced by the approach of an active oscillating region to the
innermost stable circular orbit (ISCO), a key feature of
strong-gravity general relativity (see however
\citet{Mendez:2006qq} for a different interpretation).  With the
availability of more data in the RXTE archive, it is now
possible to search for a quality factor drop similar to that
seen in other sources.

Third, \4u has recently drawn further attention, after the
detection of burst oscillations at 414 Hz with the Burst Alert
Telescope on board SWIFT \citep{Strohmayer:2008gd}. The latter
frequency which is likely to be the neutron star spin frequency
($\nu_s$) is to be compared with the frequency difference
$\Delta\nu\sim 320$ Hz between twin kHz QPOs reported so far
\citep{van-Straaten:2000zl}. Clearly the ratio $\Delta\nu/\nu_s$
was not consistent with either 0.5 or 1, observed in similar
systems. This particular result was consistent with recent suggestions by
\citet{Yin:2007cs} and \citet{Mendez:2007ph} that the kHz QPO
frequency difference may not have a strong connection to the
neutron star spin frequency in some sources (but see also the discussion in
\citealt{Barret:2008pi}). A closer inspection of the frequency difference of the
twin QPOs over a much larger data set is therefore needed in
light of this result, but also because a significant scatter
is present in the values reported by \citet{van-Straaten:2000zl}.

Here we perform an analysis of more than 11 years of RXTE data
on 4U~0614+09.  In \S~2 we describe our analysis scheme and
present our main results.  We discuss the implications of our
results in \S~3.


\section{Observations and data analysis}
\begin{figure}
   \begin{center}
   \includegraphics[width=.485\textwidth]{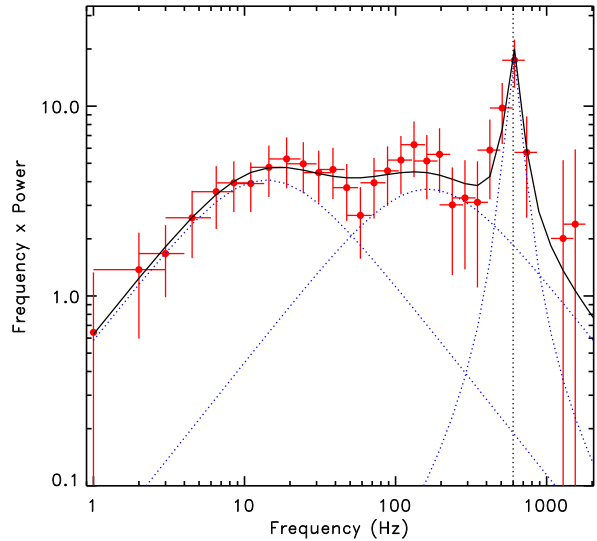}
      \caption{Illustration of the multi-lorentzian fit of the PDS for the ObsID 10095-01-01-00 recorded on April 22nd, 1996 at 22:32 pm. The broad zero-centered Lorentzian peaking around 200 Hz must be included in the fit to recover properly the parameters of the kHz QPO at 600 Hz. This was also noticed by \citet{van-Straaten:2002db}, who adjusted the low frequency continuum with one or two zero centered Lorentzians.}
   \label{boutelier:fig1}
   \end{center}
\end{figure}

We have retrieved  from the HEASARC archive science event mode data
recorded by the RXTE Proportional Counter Array (PCA). The data set
spans over eleven years from February 26th, 1996 to October 17th,
2007.  We consider segments of continuous observation (ObsIDs): 763
ObsIDs were analyzed with a typical duration of 3000 seconds. For
each ObsID, we have computed an average Power Density Spectrum (PDS)
with a 1 Hz resolution, using events recorded between 2 and 40 keV.
The PDS are normalized according to \citet{Leahy:1983mb}, so that the
Poisson noise level is constant around 2. The PDS is then blindly
searched for excess power between 300 Hz and 1400 Hz using a scanning
technique, as presented in \citet{Boirin:2000jt}. The frequency range
searched includes the highest QPO frequency reported so far
\citep{van-Straaten:2000zl}. We have also verified that no significant excesses were detected between 1400 and
2048 Hz. This justifies the use of the $1400-2048$~Hz range to estimate accurately the Poisson noise level in each observation. Each excess (at most the 2 strongest) is then fitted with a Lorentzian with three free parameters; frequency, full width at half
maximum (constrained to range from 2 to 1000 Hz), and amplitude (equal to the integrated power of the Lorentzian). The Poisson noise level is fitted separately above 1400 Hz and then frozen when fitting the QPOs. Errors on each parameter are computed with $\Delta\chi^2=1$. As in previous papers in this field, our threshold for QPOs is related to the ratio (hereafter $R$) of the Lorentzian amplitude to its $1\sigma$ error\footnote{The Lorentzian function used in the fit is $\rm Lor(\nu) = A \times W / (2 \pi) / [ (\nu-\nu_0)^2 + (W/2)^2 ]$, where A is the integrated power of the Lorentzian from 0 to $\infty$, W its width and $\nu_0$ its centroid frequency. The fitted function is linear in A, and therefore its error can be computed using $\Delta \chi^2$ (e.g. \citet{Press:1992yf}). The RMS amplitude is a derived quantity, computed as RMS=$\rm \sqrt{A/S}$, where S is the source count rate \citep{van-der-Klis:1989kn}. In this paper, we have defined $R=A/\delta A$, from which the error on the RMS is estimated as $\rm \delta RMS = 1/2 \times RMS \times R^{-1}$ after neglecting the term $\rm \delta S/S$ in the derivative of the RMS equation.} ($R$ was often quoted and used as a significance).  In this paper, our threshold for $R$ is 3, meaning that we consider only QPOs for which we can measure the power of the Lorentzian with an accuracy of $3\sigma$ or more. Such a  threshold corresponds to a  $\sim 6\sigma$ excess power in the PDS for a single trial, equivalent to $\sim 4\sigma$ if we account for the number of trials of the scanning procedure \citep{van-der-Klis:1989kn}. Furthermore, as expected we have found that $R$ positively correlates with the single trial significance of the excess power in the PDS. It is worth mentioning that the proportionality coefficient is close to 2; i.e. the QPO with the highest R ratio $\sim 10$ corresponds to a $\sim 20\sigma$ (single trial) excess power in the associated PDS. The integrated power of the Lorentzian is then converted into a root mean square (RMS), expressed as a fraction of the total source count rate. As said above the QPOs in \4u can be broad.
In order to recover properly the parameters of the QPOs in the low
frequency end ($\sim 400-700$ Hz), the continuum underneath the QPO
must be accounted for. Following \citet{van-Straaten:2002db}, the
continuum is adjusted with one or two zero centered Lorentzians. An
example of such a fit is shown in Fig. \ref{boutelier:fig1}, showing
the decomposition of the PDS into several components.

\begin{figure*}
   \begin{center}
   \includegraphics[width=.485\textwidth]{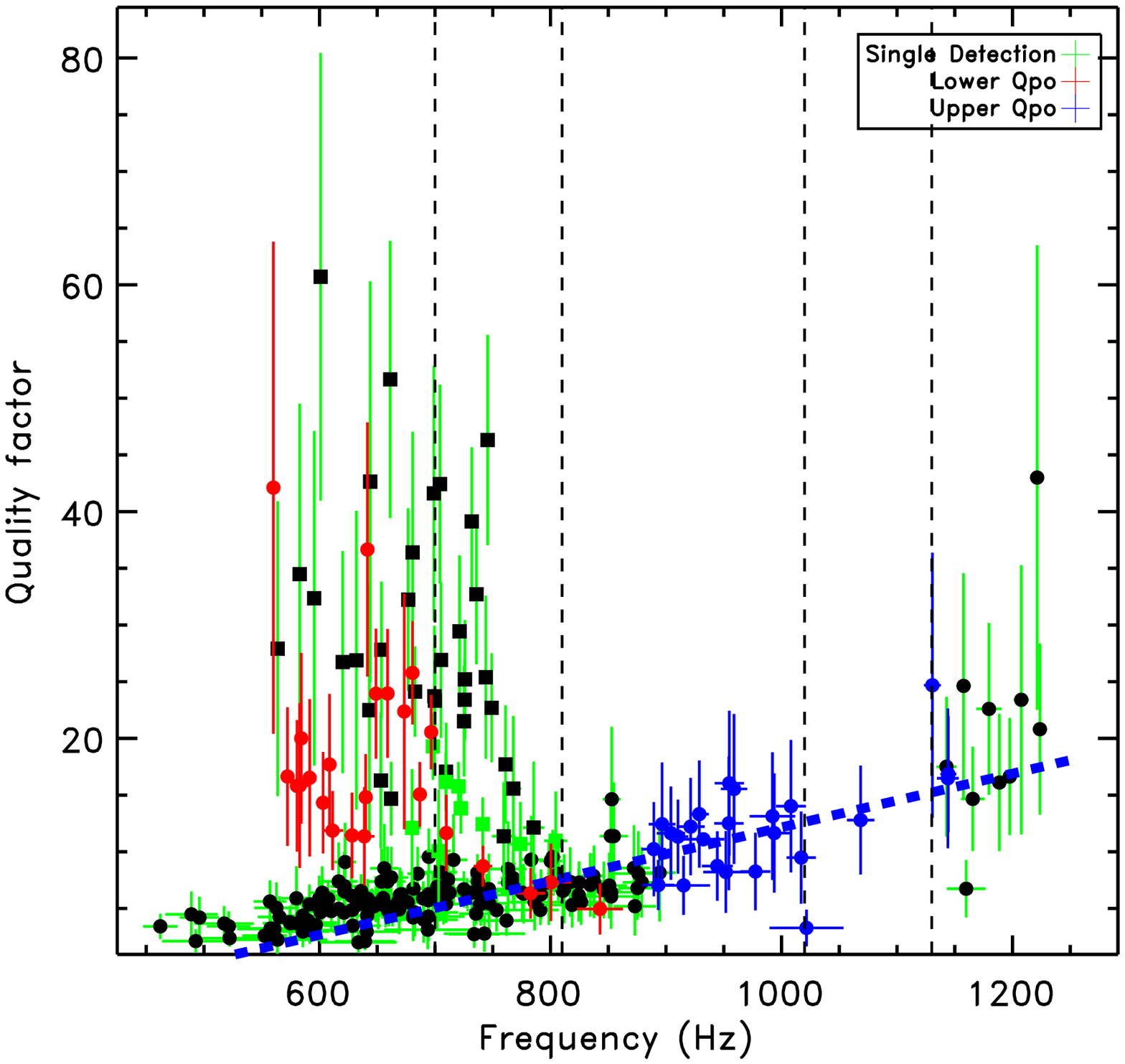} \includegraphics[width=.485\textwidth]{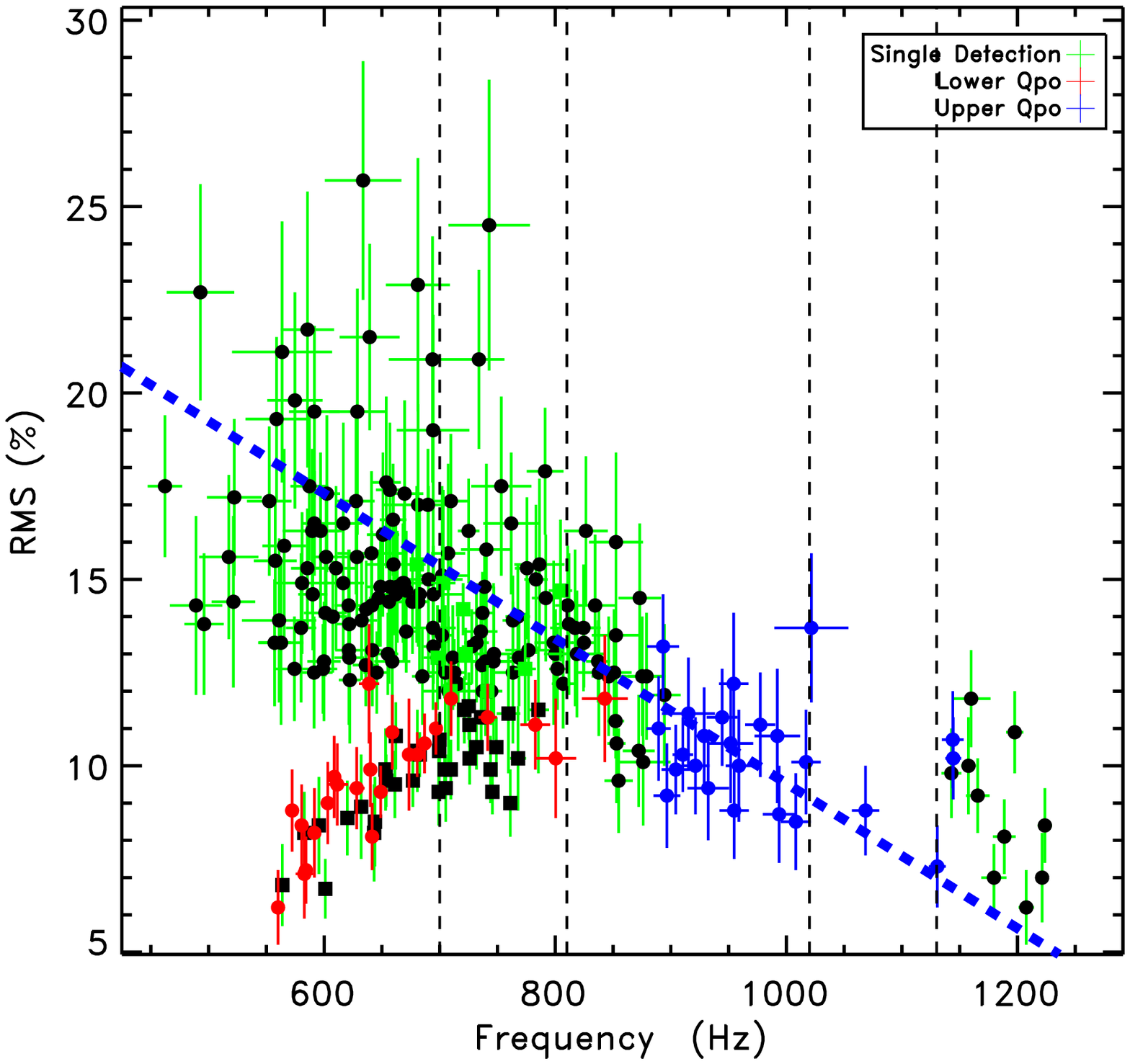}
   \caption{ Quality factor (left) and RMS amplitude (right) versus frequency of all QPOs detected in the 2--40 keV range. Each point represents the average over one ObsID. Red and blue filled circles are respectively for lower and upper twin QPOs. Black filled squares with green error bars are for single detected QPOs, identified as lower QPOs. Black filled circles with green error bars are for single detected QPOs identified as upper QPOs.}
   \label{boutelier:fig2}
   \end{center}
\end{figure*}

As a result of this systematic analysis, in 210 ObsIDs we detected a single QPO, and in 24 ObsIDs we
found two simultaneous QPOs (in one ObsID, two QPOs were
detected but not simultaneously). Two tables summarizing all the QPO detections are available in the electronic version of the paper. The remaining ObsIDs, which
contained no QPOs,
were removed from the subsequent analysis. The quality factor and RMS
amplitude versus frequency of all detected QPOs are shown in Fig
\ref{boutelier:fig2}. The patterns identified in other similar
systems by \citet{Barret:2006it} are also seen for \4u. In
particular, in the quality factor versus frequency plot, lower and
upper QPOs occupy two distinct regions: the quality factor of lower
QPOs is larger than the quality factor of upper QPOs and the quality
factor of the upper QPOs rises steadily with frequency (note that
the scatter in the quality factor of the lower QPOs is in
part because it is not corrected for the frequency drift
within the ObsID). Similarly, there is a clear trend (albeit with
some scatter) for the RMS amplitude of the upper QPOs to
decrease with increasing frequency, whereas the RMS amplitude of
the lower QPOs increases first to reach a plateau (see below for a
full description of the amplitude of the lower QPO). Note also
that the RMS amplitude of the upper QPO, when its frequency is
around 1150-1250 Hz lies above the extrapolation of the blue
dashed line, passing through the RMS values measured at lower
frequencies. Finally, it is worth mentioning that there is a gap
in the frequency distribution of the upper QPO, between 1020 and
1130~Hz (this point will be discussed in more detail below).

\subsection{On the 1330 Hz QPO}
\begin{figure}
    \begin{center}
\includegraphics[width=.45\textwidth]{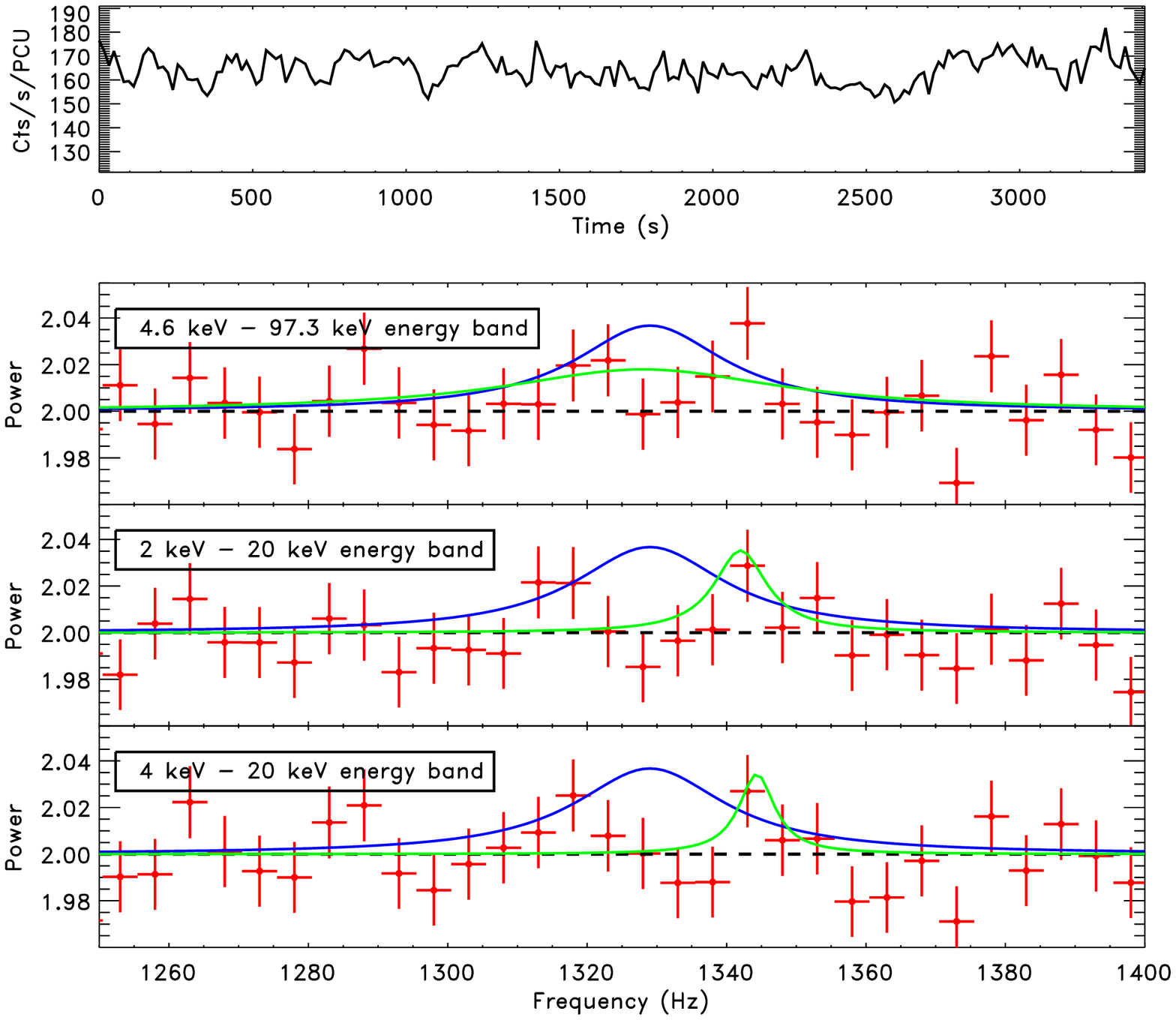}
   \caption{Top : Light curve of the ObsID 40030-01-04-00 in the energy band $4.6$~keV - $97.3$~keV. Second from the top : PDS in the energy band $4.6$~keV - $97.3$~keV with a 5 Hz resolution, as in \citet{van-Straaten:2000zl}. Third and fourth from top : PDS in the energy bands $2$~keV - $20$~keV, $4$~keV - $20$~keV with a 5 Hz resolution. The blue curve represents the $1330$~Hz QPO with parameters from \citet{van-Straaten:2000zl} and the green curve represents our best fit after initialization of the fit parameters with the values of \citet{van-Straaten:2000zl}.}
   \label{boutelier:fig3}
   \end{center}
\end{figure}

As can be seen from Fig \ref{boutelier:fig2}, the highest QPO frequency
detected in our systematic analysis is at about 1220 Hz; hence we do
not detect any QPOs at frequencies similar to the one reported at 1330
Hz by \citet{van-Straaten:2000zl}. For the 1330 Hz QPO, note that our detection threshold ($R=3$) corresponds to a single trial significance of $6\sigma$, much larger than the $3.5\sigma$ single trial significance reported by \citet{van-Straaten:2000zl}. Still, we have repeated the analysis of \citet{van-Straaten:2000zl} for the ObsID 40030-01-04-00 in which the later QPO was
reported (considering events from 5 keV to 97 keV). The strongest excess we could fit was at $\nu=1328.4\pm26.5$~Hz,
$\textrm{FWHM}=46.2\pm70.2$~Hz, $\textrm{RMS}=4.6\pm1.7$~\%, hence with a ratio $R$ of 1.3, a value far too small to claim a detection. Note that \citet{van-Straaten:2000zl} reported a larger ratio $R\sim2.8$; a value closer to our threshold, but which we failed to reproduce (see Figure \ref{boutelier:fig3}).

We have tried to optimize the energy band over which the PDS is
computed to determine whether the significance of the above excess
could be increased. By looking at the count spectrum of the source,
it dominates over the background between 2 keV and 20 keV.  We have
thus computed a PDS of this ObsID using only events from 2 to 20
keV.  The strongest excess of the PDS is no longer around 1330 Hz (a
$R\sim 1$ excess exists at 966 Hz). By initializing the
parameters of the fit with the values of
\citet{van-Straaten:2000zl}, the Lorentzian parameters are badly
constrained (as expected for a non significant excess) and
discontinuity in the $\chi^{2}$ curves prevents us from evaluating
the errors. Because it is a general property of QPOs that their RMS
amplitude increases with energy, we have also restricted the energy
range from 4 to 20 keV, but failed to detect any significant
excesses. We conclude that the 1330 Hz QPO
was likely a statistical artifact,
a possibility also implied by \citet{van-Straaten:2000zl}. A
summary plot of the analysis performed is presented in Fig
\ref{boutelier:fig3}. We also present the X-ray light curve
corresponding to the ObsID, indicating no anomalies in the source
behavior along the observation.

\subsection{Average properties of kHz QPOs}
\begin{figure*}
   \begin{center}
\includegraphics[width=.485\textwidth]{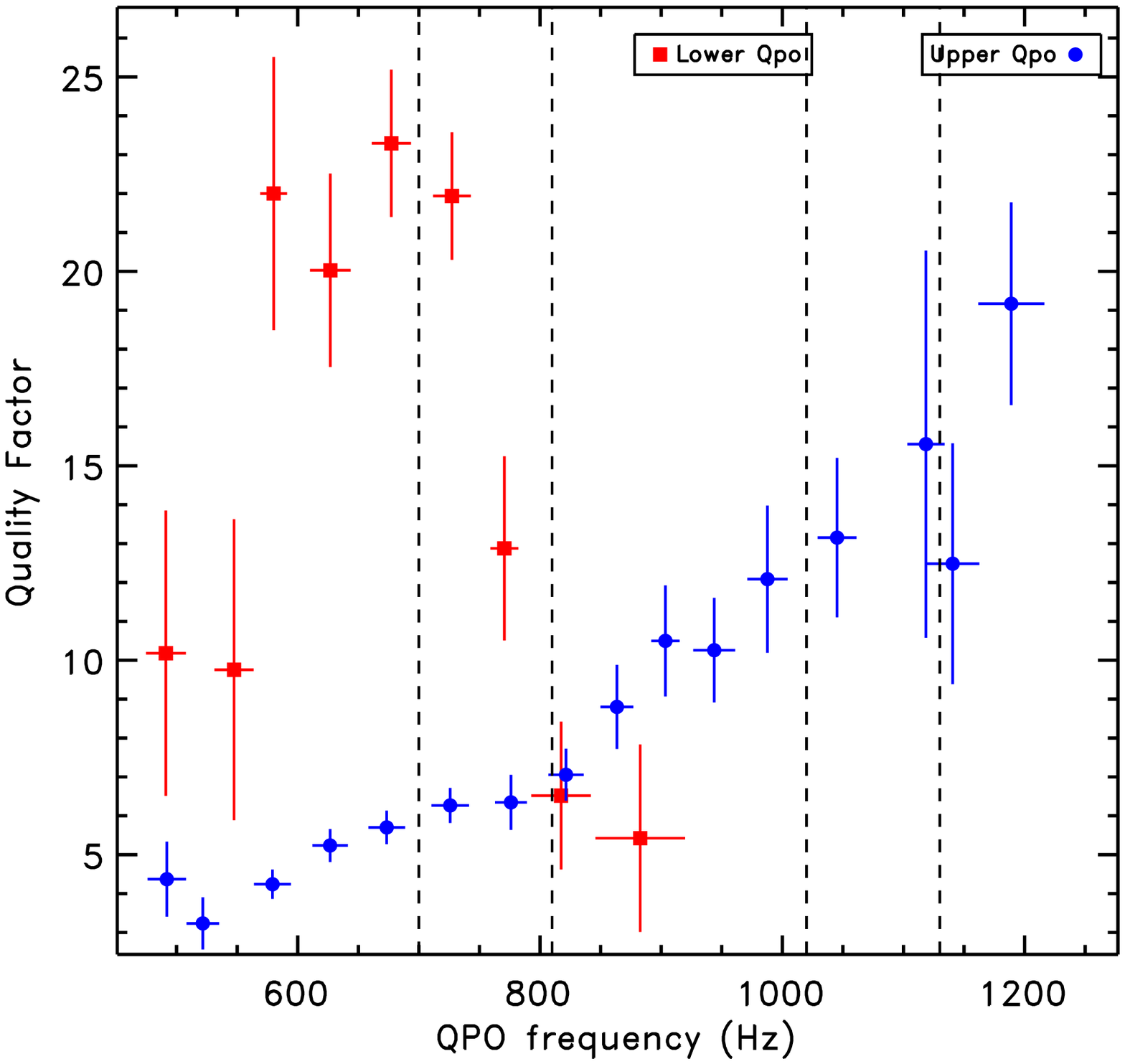} \includegraphics[width=.485\textwidth]{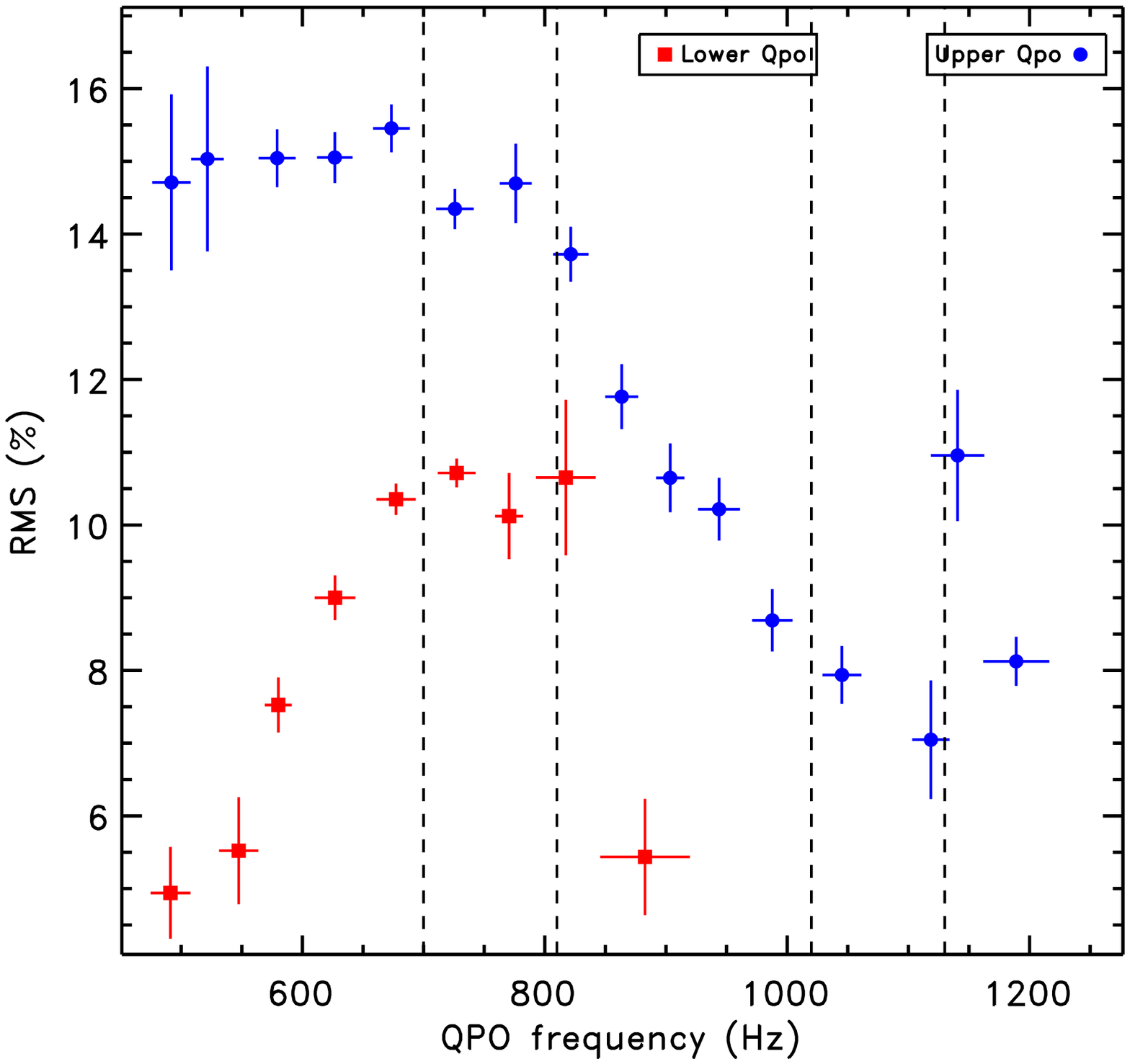}
   \caption{Quality factor (left) and RMS amplitude (right) versus frequency after grouping the ObsIDs with QPOs of similar type and frequency (within a 50 Hz interval). Red filled squares are for lower QPO. Blue filled circles are for upper QPO. An abrupt drop of coherence of the lower QPO around 700 Hz is now revealed.}
   \label{boutelier:fig4}
   \end{center}
\end{figure*}

Following on \citet{Barret:2006it}, we can identify lower and upper
QPOs based on their position in the diagrams presented in Fig
\ref{boutelier:fig2}. One can then align QPOs of similar type
(either a lower or an upper) using a shift-and-add technique
\citep{Mendez:1998hb}. This allows us to obtain a better
description of their average properties; quality factor and RMS
amplitude. We have therefore aligned all the ObsID averaged PDS, containing a QPO with
$R\ge3$ (either a lower or an upper), within a 50 Hz interval. As stated
above, our conservative threshold ensures that we do add real QPOs. Through
extensive simulations of PDS with QPO parameters appropriate for 4U0614+091
and statistics comparable to the real data, we have checked that with our
procedure, which leads to averaging a large number of 1 second PDS (several
thousands), the QPO parameters (quality factor and RMS amplitude) so
recovered are not biased by any statistical fluctuations of the signal
around the mean QPO profile. The results are presented in Fig. \ref{boutelier:fig4}. This figure shows for the first time that the quality factor of the
lower QPO starts by increasing with increasing frequency and then
drops when it reaches a frequency around 700 Hz (the last point for
the lower kHz is obtained after aligning all the PDS with an
identified upper QPO above 1100 Hz). The maximum frequency averaged value of the quality factor of the lower QPO is only about 25, while in one observation it reaches 60 albeit with large error bars. The low values reported can be explained in part by the fact that no correction for the frequency drift is applied within the ObsIDs. 
At the same time, the quality factor of the upper QPOs increases
steadily. The behavior of the RMS amplitude of the lower QPO is
consistent with other sources: it increases, saturates and then
decreases sharply with increasing frequency. On the other hand, for the
upper QPO, its RMS amplitude decreases up to a minimum around 1100
Hz, after which a second maximum is observed at 1150 Hz. Although
the minimum was not completely sampled by the data presented in
\cite{van-Straaten:2002db}, a similar trend could be inferred.

\subsection{A constant frequency difference of twin QPOs}
\begin{figure}
   \begin{center}
 	 \includegraphics[width=.485\textwidth]{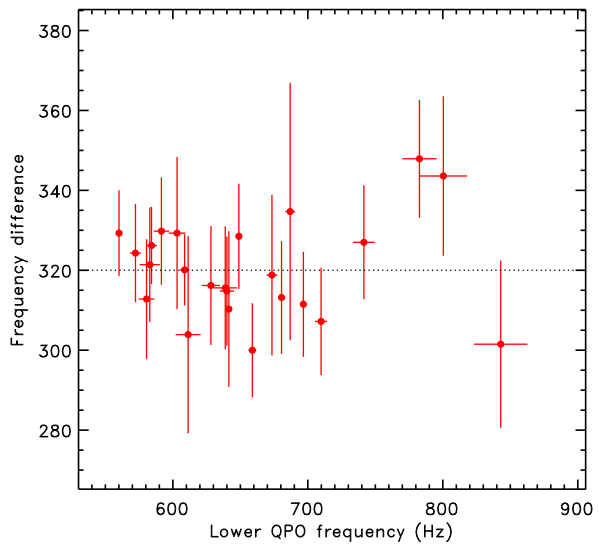}
   \caption{Frequency difference of the 24 simultaneous twin QPOs detected in our analysis. The frequency difference is consistent with being constant around 320 Hz.}
   \label{boutelier:fig5}
   \end{center}
\end{figure}

As said above, in 24 ObsIDs, we detect simultaneous twin QPOs.
Their frequency difference is plotted against the frequency of
the lower QPO in Fig \ref{boutelier:fig5}. This figure shows
that albeit with large error bars, the frequency difference is
consistent with being constant around 320 Hz. A similar
conclusion was reached by \citet{van-Straaten:2000zl} using a
reduced data set, but with a rather large scatter, with
values ranging from 250 and 380 Hz. This value is indeed
significantly different from the neutron star spin frequency
at 414 Hz \citep{Strohmayer:2008gd}. Note also that the drop
towards higher frequencies seen in other sources is not
observed in \4u, and our result is consistent with findings of
\citet{Ford:1997hl} and \citet{van-Straaten:2000zl}.

\subsection{On the distribution of twin QPO frequencies}
We have started to investigate how the dependence of the QPO
quality factor (or width) and RMS amplitude with frequency
influences their detectability in time intervals of fixed
durations, hence affects the observed distribution of
frequency ratios. This is an important issue, because such
distributions have been shown to be peaked around small
integer ratios in some sources, e.g. 3:2 in the case of Sco X-1
\citep{Abramowicz:2003sf} and 4U1636-536
\citep{Torok:2008lr}. This has been used as an argument in
favor of resonance based models (but see the discussion in
\cite{Belloni:2005vl}). Following this idea,
\citet{2008NewAR..51..835B} have shown that in the case of
4U1820-303, a gap in the frequency distribution was present,
together with a cluster of frequency ratio. They showed that
the lack of twin QPOs within the gap could not be due to a
lack of sensitivity for QPO detection, provided that the
parameters of the QPOs (RMS and width) could be interpolated
within the gap, using values measured before and after.
Their result implied a sudden change of the QPO properties
within the gap, most likely a loss of coherence of the upper
QPO.

\begin{figure}
   \begin{center}
 	 \includegraphics[width=.485\textwidth]{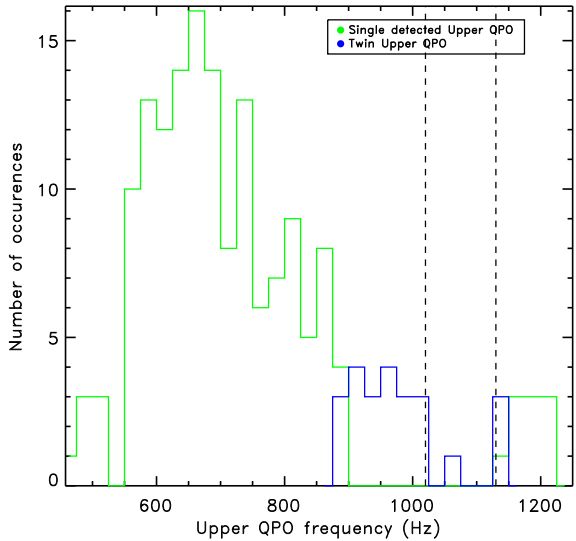}
   \caption{Histograms of occurrence of single and twin upper QPOs showing a frequency gap between 1020 Hz and 1130 Hz in the distribution of frequencies. A frequency bin of 20 Hz has been used. The gap is indicated for twin upper QPOs as vertical dashed lines.}
   \label{boutelier:fig6}
   \end{center}
\end{figure}

As said above, there is a gap in the frequency distribution of the
twin upper QPOs, between 1020 Hz and 1130 Hz. The histograms of occurrence of single and twin upper kHz QPOs are shown in Figure \ref{boutelier:fig6}.

Between 1020 Hz and 1130 Hz, there is only one detection of a twin upper QPO and no single
upper QPO detected. It is possible to evaluate the probability of
having such a gap of 110 Hz width, with one QPO inside, assuming
that the 24 twin upper QPO frequencies are uniformly distributed over
their 250 Hz frequency span. This probability is less than $3.7
\times 10^{-4}$ ($3.4\sigma$), giving us confidence that the gap
is significant. As shown in Figure \ref{boutelier:fig6}, single upper kHz QPOs are detected over a much wider range of frequencies and show a highly non uniform and peaked distribution of frequencies; single upper kHz QPOs are detected predominantly below the gap, but some are also detected above.

There is no such a gap in the frequency distribution of the lower
QPOs, 320 Hz below (there are more than 10 single lower QPOs).  In
fact, it is the frequency range in which both the quality factor
and the amplitude of the lower QPO reaches a maximum; hence where
lower QPOs are easy to detect. After recovering the upper QPO
parameters within the gap through the shift-and-add technique,
Figure \ref{boutelier:fig4} shows that the gap corresponds to the
region where the RMS amplitude of the upper QPO reaches a minimum.
This leads to the conclusion that the absence of twin QPOs within
the gap is linked to a localized drop of the amplitude of the
upper QPO, unlike in 4U1820-303 for which the gap was more likely
related to a local minimum of the quality factor of the upper QPO. 
This result suggests some cautions when estimating frequency
ratios based on the detection of single lower QPOs\footnote{and
computing the frequency of the upper QPO through the linear
function that links the two QPO
frequencies.}\citep{Belloni:2005vl,Belloni:2007ad}. Clearly in sensitivity limited observations (as is the case here, where the data are analyzed over comparable integration times), the histogram of ratios computed from simultaneous twin QPOs will differ from the one computed from the distribution of single lower QPOs. The study of
the histogram of frequencies and frequency ratios, resulting from
the frequency dependency of the QPO RMS and width (see Figure
\ref{boutelier:fig4}), will
be the subject of a forthcoming paper.

\section{Conclusions}
The main results of our systematic analysis of all archival RXTE data for
\4u are:
\begin{itemize}
\item We do not confirm the previous claim of a QPO at 1330 Hz.
This is based on a thorough reanalysis of the observation from
which the QPO was reported, and the fact that in our analysis,
the highest frequency detected is at $\sim 1220$ Hz.  This value is
fully consistent with maximum frequencies observed in similar
systems.
\item We observe for the first time a drop of the quality factor of
the lower QPO. Such a drop has been interpreted as being related to
the oscillating region, crossing the innermost stable circular
orbit, and our detection is consistent with that idea.
\item The frequency difference between the lower and upper kHz QPOs is consistent with being constant, around 320 Hz, a value significantly different from the neutron star spin frequency (414 Hz) or half its value.
\item If the drop in quality factor for the lower kHz QPO is due to the ISCO,
one can estimate the orbital frequency there, given the constant frequency difference.
To do this we compute the maximum frequency of the lower QPO by
extrapolating its quality factor to zero. This yields a limiting
frequency of $\sim 920-930$ Hz, corresponding to an orbital frequency
of 1250 Hz at the innermost stable orbit ($\nu_{ISCO}$). As a safety
check, we note that the maximum QPO frequency detected for the upper
QPO is indeed lower than $\nu_{ISCO}$. From this we can estimate the
mass of the NS following the equation $\frac{M}{M_{\odot}} \approx
{2200\textrm{~Hz}}\times (1 + 0.75j)/{\nu_{ISCO}}$ where $j=
\frac{cJ}{GM^{2}} \sim 0.1 - 0.2$ is the dimensionless angular
momentum of the star. This leads to a gravitational mass for the
neutron star of 1.9 $M_{\odot}$, i.e. a relatively massive NS, but
still consistent with realistic modern equations of state, which
predict maximum masses for slowly rotating  ( $j \ll 1$) stars of
$\sim 1.8-2.3~M_{\odot}$ \citep{Akmal:1998hb,Lattimer:2001vf,Klahn:2006uk}.
\item \4u is the second object in which a gap in the frequency distribution of the upper QPO is
observed, after 4U1820-303. This gap is likely associated
with a local minimum of the RMS amplitude of the upper QPO. This
is the first time such a minimum is so clearly identified,
although it was suggested in previous work
\citep{van-Straaten:2002db}. Similar trends will be searched in
similar systems, because frequency gaps are present in the data
analyzed by \cite{Barret:2006it} in sources like 4U1636-536.
This result motivates caution when estimating
frequency ratios based on the detection of single lower QPOs.
\end{itemize}
We note two additional consequences of our work.  First,
our conclusion that the proposed 1330~Hz QPO in
\4u is not likely to be real means that no kHz QPO
source has a confirmed frequency significantly above
$\sim 1200$~Hz.  This is surprising a priori, because
for a neutron star of canonical mass $M=1.4~M_\odot$
and a spin parameter of $j=0.1$, the orbital frequency
at the ISCO is 1690~Hz.  Such frequencies should have been detected with RXTE, hence the wide gap relative
to the maximum actually observed suggests a physical
cause.  One possibility is that these stars tend to
have higher masses, consistent with accretion of several
tenths of a solar mass, but other explanations should
also be explored.

Second, \4u is an important source to test the suggestion
by \cite{2005MNRAS.361..855B,Barret:2006it,2007MNRAS.376.1139B} that the rapid drop in coherence
of the lower QPO is caused by the approach of the
oscillating region to the ISCO.  In their proposal, the
orbital frequency at the ISCO can be very roughly
estimated by adding the separation frequency $\Delta\nu$
between the twin kHz QPOs to the maximum frequency seen
from the lower QPO after a sharp drop in quality factor.
If the 1330~Hz QPO were real, it would contradict this
interpretation because the maximum inferred lower
frequency is $\sim 900$~Hz and the separation is $\sim
320$~Hz. However, the maximum significant upper QPO
frequency is actually $\sim 1220$~Hz, which is consistent
with the ISCO interpretation.  This is encouraging, but
other sources must also be analyzed carefully to look for
potential disproofs of our hypothesis.
\section{Acknowledgements}
MCM was supported in part by US NSF grant AST0708424.  We are grateful to the referee, Michiel van der Klis for very helpful comments, and for double checking and confirming in his report that the 1330 Hz QPO previously claimed cannot be reproduced with the archived data. We are also thankful to Mariano Mendez for many exchanges on data analysis, in particular about the way to assess the significance of kHz QPOs.

\section{Tables}
\begin{table*}
\caption{The parameters of the single detected QPOs together with information concerning the observations: the ObsID name (ObsID), the date of the observation (Date), the start time of the observation (Time, UTC), the observation duration (T$_{\rm obs}$), the source count rate divided by the number of active PCU units (Rate), the background count rate (also divided by the number of active PCU units) (Bkg), the frequency ($\nu$), Full Width Half Maximum (FWHM), fractional RMS amplitude (RMS, \%) and the $R$ ratio, which is computed as the Lorentzian integrated power divided by its $1\sigma$ error (the larger R, the more significant the QPO: R$\sim3$ corresponds to a $\sim 6\sigma$ (single trial) excess in the power spectrum, see text for details).}
\begin{center}
\begin{tabular}{cccccccccccc}
\hline
\hline
ObsID & Date & Time & T$_{\rm obs}$ & Rate & Bkg & $\nu$ & FWHM & RMS & R \\
\hline
\hline
10072-02-02-00&03-16-1996&09:10&$3456.0$&$92.8$&$19.1$&$725.0\pm9.3$&$108.3\pm24.4$&$16.3\pm1.4$&$6.0$\\
10072-02-02-00&03-16-1996&10:46&$1120.0$&$95.4$&$19.1$&$753.3\pm26.0$&$154.9\pm55.1$&$17.5\pm2.4$&$3.6$\\
10095-01-01-00&04-22-1996&20:56&$3440.0$&$89.0$&$17.9$&$565.4\pm18.8$&$133.1\pm52.8$&$15.9\pm2.6$&$3.1$\\
10095-01-01-00&04-22-1996&22:32&$3456.0$&$108.5$&$17.9$&$601.0\pm18.4$&$128.6\pm37.6$&$14.1\pm1.7$&$4.2$\\
10095-01-02-00&04-24-1996&13:19&$2192.0$&$133.6$&$20.9$&$582.8\pm2.1$&$16.9\pm7.3$&$8.2\pm1.1$&$3.7$\\
10095-01-02-00&04-24-1996&18:11&$1952.0$&$160.6$&$20.9$&$894.4\pm14.1$&$109.7\pm56.1$&$11.9\pm1.9$&$3.1$\\
10095-01-02-00&04-24-1996&20:58&$3456.0$&$151.5$&$20.9$&$852.2\pm7.3$&$74.8\pm18.8$&$11.2\pm1.0$&$5.5$\\
10095-01-02-00&04-24-1996&22:34&$3456.0$&$143.0$&$20.9$&$801.9\pm7.2$&$84.1\pm20.4$&$12.6\pm1.1$&$5.8$\\
10095-01-02-00&04-25-1996&00:10&$3456.0$&$138.3$&$20.9$&$763.3\pm8.8$&$105.0\pm24.6$&$13.9\pm1.2$&$6.0$\\
10095-01-02-00&04-25-1996&01:46&$3456.0$&$144.9$&$20.9$&$806.8\pm9.7$&$100.1\pm28.3$&$12.2\pm1.2$&$5.0$\\
10095-01-02-00&04-25-1996&03:22&$3456.0$&$147.6$&$20.9$&$824.8\pm10.4$&$113.2\pm27.4$&$13.3\pm1.2$&$5.8$\\
10095-01-03-00&08-07-1996&03:30&$2672.0$&$110.3$&$20.0$&$837.7\pm18.2$&$114.0\pm47.1$&$12.5\pm1.7$&$3.6$\\
10095-01-03-010&08-07-1996&19:18&$3456.0$&$79.2$&$17.7$&$587.2\pm16.3$&$144.5\pm41.9$&$17.5\pm2.0$&$4.4$\\
20064-16-01-01&08-15-1997&00:02&$3376.0$&$117.3$&$19.4$&$636.0\pm11.2$&$97.3\pm20.3$&$12.7\pm1.1$&$5.7$\\
20074-01-01-00&01-10-1997&10:15&$3408.0$&$99.5$&$17.9$&$585.5\pm14.0$&$134.3\pm44.4$&$15.3\pm1.6$&$4.8$\\
20074-01-02-00&01-25-1997&13:37&$3424.0$&$138.5$&$21.1$&$641.7\pm8.3$&$109.6\pm30.9$&$13.1\pm1.5$&$4.3$\\
20074-01-03-00&03-16-1997&22:23&$2960.0$&$124.4$&$17.9$&$663.3\pm11.8$&$141.3\pm35.8$&$14.8\pm1.5$&$5.1$\\
20074-01-04-00&05-07-1997&10:12&$2240.0$&$158.9$&$20.5$&$739.3\pm12.0$&$109.1\pm31.0$&$11.8\pm1.2$&$4.9$\\
20074-01-05-00&05-16-1997&05:22&$3456.0$&$138.6$&$20.0$&$655.1\pm8.9$&$110.7\pm22.9$&$13.0\pm1.1$&$6.0$\\
20074-01-06-00&08-18-1997&22:28&$3440.0$&$132.3$&$20.1$&$574.3\pm17.4$&$151.1\pm37.8$&$12.6\pm1.4$&$4.6$\\
20074-01-07-00&09-03-1997&21:10&$2976.0$&$126.6$&$19.0$&$636.4\pm9.0$&$117.5\pm29.7$&$14.2\pm1.3$&$5.4$\\
20074-01-08-00&09-06-1997&17:52&$2848.0$&$126.6$&$19.8$&$621.2\pm12.1$&$134.2\pm34.5$&$14.3\pm1.5$&$4.7$\\
20074-01-10-00&10-18-1997&18:19&$3456.0$&$109.1$&$20.4$&$552.5\pm19.2$&$212.2\pm56.9$&$17.1\pm2.0$&$4.3$\\
30053-01-01-01&10-20-1998&09:04&$2496.0$&$83.9$&$17.6$&$610.3\pm18.8$&$127.9\pm46.1$&$15.3\pm2.2$&$3.5$\\
30053-01-01-02&10-20-1998&13:57&$2256.0$&$85.0$&$16.9$&$650.9\pm13.9$&$120.7\pm46.5$&$16.2\pm2.2$&$3.7$\\
30053-01-01-02&10-20-1998&15:28&$2976.0$&$86.7$&$16.9$&$657.2\pm16.8$&$126.8\pm42.0$&$14.8\pm1.9$&$4.0$\\
30053-01-01-02&10-20-1998&17:04&$3408.0$&$89.8$&$16.9$&$671.0\pm11.2$&$107.2\pm32.9$&$14.7\pm1.6$&$4.6$\\
30053-01-02-00&10-16-1998&06:04&$2800.0$&$133.6$&$17.9$&$709.3\pm3.3$&$43.9\pm6.8$&$12.0\pm0.7$&$8.5$\\
30053-01-02-01&10-16-1998&02:55&$2576.0$&$128.9$&$17.8$&$653.6\pm1.9$&$23.5\pm5.0$&$9.6\pm0.7$&$6.5$\\
30053-01-02-02&10-16-1998&18:40&$3520.0$&$133.9$&$16.8$&$682.7\pm1.8$&$28.3\pm4.6$&$10.3\pm0.6$&$8.6$\\
30053-01-02-05&10-20-1998&12:17&$2448.0$&$85.1$&$17.3$&$653.7\pm12.5$&$88.7\pm33.3$&$17.6\pm2.3$&$3.8$\\
30054-01-01-00&05-18-1998&03:58&$3424.0$&$166.0$&$21.1$&$781.8\pm7.2$&$82.4\pm17.7$&$10.4\pm0.8$&$6.3$\\
30054-01-01-00&05-18-1998&05:34&$3456.0$&$159.0$&$21.1$&$736.7\pm5.4$&$90.4\pm17.6$&$12.7\pm0.9$&$6.8$\\
30054-01-01-02&05-18-1998&08:47&$3376.0$&$157.5$&$18.3$&$744.6\pm9.6$&$113.3\pm24.1$&$12.0\pm0.9$&$6.5$\\
30054-01-01-02&05-18-1998&10:23&$3392.0$&$155.4$&$18.3$&$756.2\pm7.5$&$97.6\pm21.1$&$11.6\pm0.9$&$6.5$\\
30054-01-01-02&05-18-1998&11:59&$1776.0$&$155.7$&$18.3$&$737.0\pm9.1$&$112.5\pm22.6$&$14.1\pm1.1$&$6.3$\\
30054-01-01-03&05-19-1998&03:58&$3456.0$&$151.4$&$21.6$&$712.6\pm9.4$&$111.7\pm21.0$&$12.5\pm0.9$&$6.8$\\
30054-01-01-03&05-19-1998&05:34&$1776.0$&$154.9$&$21.6$&$731.8\pm10.6$&$114.7\pm36.5$&$13.3\pm1.6$&$4.3$\\
30054-01-01-04&05-19-1998&08:46&$3440.0$&$142.6$&$18.2$&$671.0\pm8.6$&$120.7\pm23.6$&$13.6\pm1.1$&$6.4$\\
30054-01-01-04&05-19-1998&10:22&$2912.0$&$139.7$&$18.2$&$647.1\pm8.3$&$124.7\pm27.6$&$14.5\pm1.3$&$5.8$\\
30054-01-01-06&05-17-1998&03:58&$3440.0$&$133.0$&$21.1$&$591.5\pm9.8$&$159.2\pm32.6$&$16.5\pm1.4$&$6.0$\\
30054-01-01-06&05-17-1998&05:34&$3456.0$&$137.8$&$21.1$&$645.5\pm11.4$&$117.6\pm25.4$&$12.5\pm1.1$&$5.9$\\
30054-01-02-00&05-22-1998&02:23&$3184.0$&$148.7$&$20.2$&$727.8\pm8.6$&$114.8\pm22.7$&$13.2\pm1.0$&$6.8$\\
30054-01-02-00&05-22-1998&03:59&$3456.0$&$151.6$&$20.2$&$738.8\pm8.3$&$138.1\pm22.5$&$14.8\pm0.9$&$8.1$\\
30054-01-02-00&05-22-1998&05:35&$3456.0$&$149.9$&$20.2$&$737.2\pm7.8$&$100.0\pm21.0$&$12.0\pm0.9$&$6.4$\\
30054-01-02-01&05-24-1998&02:23&$3248.0$&$134.1$&$19.9$&$599.0\pm12.2$&$116.6\pm27.9$&$12.6\pm1.2$&$5.2$\\
30054-01-02-01&05-24-1998&03:59&$3456.0$&$124.1$&$19.9$&$590.1\pm12.5$&$150.4\pm37.4$&$14.6\pm1.5$&$4.8$\\
30054-01-02-02&05-25-1998&00:47&$2880.0$&$147.3$&$20.1$&$694.5\pm8.6$&$111.1\pm22.6$&$13.2\pm1.0$&$6.4$\\
30054-01-02-02&05-25-1998&02:23&$3264.0$&$149.9$&$20.1$&$735.6\pm14.4$&$162.1\pm30.8$&$13.6\pm1.1$&$6.3$\\
30054-01-02-03&05-26-1998&02:23&$3328.0$&$141.8$&$19.5$&$743.6\pm8.8$&$92.7\pm22.3$&$11.4\pm1.0$&$5.7$\\
30054-01-02-03&05-26-1998&03:59&$3456.0$&$117.3$&$19.5$&$717.5\pm9.2$&$75.8\pm20.5$&$11.0\pm1.1$&$4.8$\\
\hline
\hline
\end{tabular}
\end{center}
\end{table*}
\setcounter{table}{0}
\begin{table*}
\caption{Continued}
\begin{center}
\begin{tabular}{cccccccccccc}
\hline
\hline
30056-01-01-00&03-10-1998&08:23&$3408.0$&$152.0$&$17.4$&$1159.9\pm16.8$&$172.1\pm62.1$&$11.8\pm1.3$&$4.5$\\
30056-01-01-00&03-10-1998&10:02&$3216.0$&$149.7$&$17.4$&$1165.6\pm10.2$&$79.5\pm24.2$&$9.2\pm1.0$&$4.5$\\
30056-01-01-00&03-10-1998&11:44&$2864.0$&$142.1$&$17.4$&$761.1\pm5.0$&$43.0\pm12.4$&$9.0\pm0.9$&$5.0$\\
30056-01-01-00&03-10-1998&13:23&$1952.0$&$134.3$&$17.4$&$705.2\pm2.6$&$26.2\pm6.6$&$9.4\pm0.9$&$5.6$\\
30056-01-01-02&03-07-1998&11:43&$688.0$&$130.8$&$17.4$&$643.9\pm2.5$&$15.1\pm6.2$&$8.5\pm1.2$&$3.4$\\
30056-01-01-04&03-09-1998&10:02&$3216.0$&$140.1$&$17.3$&$749.0\pm2.0$&$33.0\pm6.9$&$10.5\pm0.8$&$7.0$\\
30056-01-01-05&03-13-1998&11:43&$2976.0$&$85.1$&$18.0$&$689.8\pm12.2$&$117.2\pm27.9$&$17.0\pm1.5$&$5.5$\\
30056-01-01-06&03-14-1998&11:43&$3024.0$&$102.8$&$21.6$&$699.5\pm3.3$&$29.5\pm7.6$&$10.4\pm1.0$&$5.2$\\
30056-01-01-06&03-14-1998&13:24&$816.0$&$130.7$&$21.6$&$680.6\pm2.3$&$18.7\pm5.4$&$10.4\pm1.1$&$4.7$\\
30056-01-01-07&03-15-1998&11:44&$2960.0$&$90.5$&$21.1$&$707.6\pm22.5$&$104.8\pm44.4$&$12.0\pm2.0$&$3.0$\\
30056-01-02-00&03-17-1998&11:43&$3072.0$&$77.0$&$21.5$&$489.1\pm22.6$&$109.1\pm43.9$&$14.3\pm2.4$&$3.0$\\
30056-01-02-01&03-18-1998&13:25&$2720.0$&$91.5$&$21.2$&$709.6\pm14.5$&$131.5\pm37.7$&$17.1\pm1.8$&$4.8$\\
30056-01-02-02&03-19-1998&09:06&$1040.0$&$92.8$&$18.8$&$740.4\pm21.9$&$113.6\pm43.5$&$15.8\pm2.3$&$3.4$\\
30056-01-02-02&03-19-1998&10:04&$3344.0$&$101.9$&$18.8$&$800.2\pm9.8$&$87.1\pm21.4$&$13.0\pm1.2$&$5.3$\\
30056-01-02-04&03-22-1998&08:27&$3456.0$&$89.9$&$17.7$&$621.7\pm12.0$&$89.5\pm35.7$&$13.8\pm2.0$&$3.5$\\
30056-01-03-02&11-01-1998&17:00&$944.0$&$153.5$&$19.2$&$699.0\pm1.7$&$16.8\pm4.5$&$9.3\pm0.9$&$5.3$\\
30056-01-04-00&10-30-1998&02:37&$3152.0$&$122.5$&$18.1$&$837.1\pm10.4$&$106.3\pm25.2$&$12.8\pm1.1$&$5.7$\\
30056-01-04-01&10-30-1998&13:57&$2272.0$&$160.9$&$19.9$&$725.7\pm1.9$&$31.0\pm4.7$&$11.1\pm0.6$&$9.1$\\
30056-01-05-00&11-06-1998&15:23&$3312.0$&$195.0$&$18.5$&$1223.7\pm6.1$&$58.8\pm21.0$&$8.4\pm1.0$&$4.3$\\
30056-01-05-03&11-09-1998&12:24&$1728.0$&$159.1$&$18.6$&$731.8\pm1.1$&$18.7\pm3.1$&$10.5\pm0.6$&$8.8$\\
30056-01-05-04&11-12-1998&13:46&$2832.0$&$100.6$&$17.8$&$711.3\pm13.0$&$94.2\pm27.2$&$12.9\pm1.4$&$4.5$\\
40025-01-01-00&05-23-1999&00:22&$2720.0$&$115.8$&$17.2$&$658.9\pm15.6$&$89.3\pm35.2$&$12.8\pm1.9$&$3.3$\\
40025-01-01-01&05-25-1999&05:07&$3456.0$&$117.5$&$18.9$&$596.9\pm15.9$&$168.2\pm53.4$&$16.3\pm2.1$&$3.9$\\
40030-01-01-00&12-23-1998&21:32&$3456.0$&$89.8$&$18.1$&$557.1\pm14.2$&$99.2\pm30.1$&$13.3\pm1.7$&$4.0$\\
40030-01-02-00&12-24-1998&21:34&$3344.0$&$119.2$&$17.9$&$768.6\pm9.7$&$110.1\pm31.3$&$12.9\pm1.3$&$5.1$\\
40030-01-03-00&12-25-1998&21:33&$3296.0$&$207.2$&$18.0$&$1207.6\pm6.7$&$51.6\pm25.9$&$6.2\pm1.0$&$3.3$\\
40030-01-07-00&12-29-1998&23:48&$944.0$&$144.1$&$17.8$&$632.0\pm3.0$&$23.5\pm11.4$&$8.9\pm1.4$&$3.2$\\
40030-01-08-00&12-30-1998&22:05&$1264.0$&$81.4$&$17.0$&$681.3\pm30.3$&$154.4\pm64.6$&$17.0\pm2.7$&$3.2$\\
40030-01-08-00&12-30-1998&23:04&$576.0$&$80.9$&$17.0$&$681.1\pm27.7$&$161.9\pm73.9$&$22.9\pm3.4$&$3.3$\\
40030-01-09-00&12-31-1998&23:47&$816.0$&$116.9$&$17.2$&$817.1\pm21.5$&$111.3\pm43.3$&$13.7\pm2.1$&$3.3$\\
40030-01-11-00&01-02-1999&23:46&$960.0$&$91.9$&$17.8$&$627.6\pm16.1$&$102.5\pm39.3$&$17.1\pm2.5$&$3.4$\\
40030-01-13-00&01-04-1999&23:43&$1088.0$&$107.7$&$18.7$&$786.3\pm21.6$&$141.4\pm59.4$&$15.4\pm2.3$&$3.4$\\
40030-01-15-00&01-07-1999&01:23&$864.0$&$151.9$&$19.8$&$721.3\pm2.1$&$24.5\pm5.5$&$11.5\pm0.9$&$6.3$\\
40429-01-01-02&09-05-1999&17:07&$2608.0$&$172.8$&$16.7$&$739.7\pm23.3$&$119.5\pm48.3$&$12.8\pm2.0$&$3.3$\\
40429-01-01-03&09-06-1999&17:05&$2672.0$&$230.3$&$17.1$&$620.1\pm2.7$&$23.2\pm8.4$&$8.6\pm1.0$&$4.1$\\
40429-01-01-03&09-06-1999&18:41&$3040.0$&$238.7$&$17.1$&$653.2\pm3.9$&$40.1\pm14.5$&$9.9\pm1.1$&$4.3$\\
40429-01-01-03&09-06-1999&20:17&$3456.0$&$258.7$&$17.1$&$725.9\pm2.0$&$28.8\pm5.9$&$10.2\pm0.7$&$7.0$\\
40429-01-01-05&09-08-1999&18:37&$3104.0$&$205.9$&$17.4$&$873.1\pm18.3$&$167.9\pm79.8$&$14.5\pm2.0$&$3.6$\\
40429-01-01-05&09-08-1999&20:13&$3456.0$&$213.2$&$17.4$&$875.3\pm16.7$&$128.7\pm50.7$&$12.4\pm1.6$&$3.8$\\
40429-01-02-00&09-12-1999&20:05&$3504.0$&$105.1$&$18.1$&$680.2\pm7.5$&$56.2\pm23.4$&$15.4\pm2.1$&$3.6$\\
40429-01-03-06&10-21-1999&20:32&$3392.0$&$143.7$&$19.3$&$1197.6\pm7.5$&$72.0\pm21.9$&$10.9\pm1.1$&$4.8$\\
40429-01-07-02&02-09-2000&09:27&$3328.0$&$140.5$&$18.5$&$1157.6\pm6.4$&$47.0\pm18.7$&$10.0\pm1.3$&$3.7$\\
50026-02-01-000&03-13-2001&04:12&$3440.0$&$77.8$&$18.0$&$558.8\pm26.8$&$235.0\pm78.1$&$19.3\pm2.2$&$4.3$\\
50026-02-01-000&03-13-2001&07:23&$3456.0$&$67.7$&$18.0$&$492.9\pm29.2$&$232.0\pm99.1$&$22.7\pm2.9$&$3.9$\\
60037-04-01-000&12-13-2001&14:14&$3456.0$&$81.0$&$19.7$&$574.7\pm23.9$&$155.5\pm57.7$&$19.8\pm2.9$&$3.4$\\
60037-04-01-01&12-13-2001&01:32&$3456.0$&$63.7$&$17.3$&$628.7\pm24.5$&$180.7\pm74.0$&$19.5\pm3.3$&$3.0$\\
60037-04-01-01&12-13-2001&04:43&$3392.0$&$63.3$&$17.3$&$585.6\pm23.0$&$197.8\pm84.6$&$21.7\pm3.7$&$3.0$\\
60037-04-01-01&12-13-2001&06:18&$3344.0$&$63.7$&$17.3$&$602.4\pm11.4$&$94.3\pm32.6$&$17.3\pm2.1$&$4.1$\\
60424-01-01-00&12-05-2001&12:35&$3392.0$&$139.6$&$17.5$&$872.2\pm13.8$&$101.8\pm43.3$&$10.4\pm1.5$&$3.5$\\
60424-01-01-00&12-05-2001&14:10&$3392.0$&$131.6$&$17.5$&$846.2\pm14.7$&$127.5\pm45.7$&$12.4\pm1.6$&$3.8$\\
80037-01-01-00&11-03-2003&17:07&$3504.0$&$88.9$&$15.9$&$601.7\pm10.2$&$108.2\pm31.6$&$15.6\pm1.7$&$4.5$\\
80037-01-01-00&11-03-2003&18:42&$3520.0$&$88.2$&$15.9$&$561.0\pm32.1$&$175.9\pm61.7$&$13.9\pm2.2$&$3.1$\\
80037-01-01-00&11-03-2003&20:16&$3520.0$&$89.3$&$15.9$&$580.0\pm13.1$&$98.3\pm44.8$&$13.7\pm2.0$&$3.4$\\
80037-01-01-00&11-03-2003&21:51&$3520.0$&$67.9$&$15.9$&$591.5\pm21.9$&$190.9\pm67.3$&$19.5\pm2.3$&$4.3$\\
80037-01-01-01&11-04-2003&15:35&$2144.0$&$97.7$&$16.1$&$639.4\pm25.9$&$303.8\pm99.0$&$21.5\pm2.5$&$4.2$\\
80037-01-01-01&11-04-2003&16:46&$3520.0$&$90.9$&$16.1$&$648.8\pm12.7$&$120.6\pm33.3$&$14.8\pm1.6$&$4.5$\\
80037-01-01-01&11-04-2003&18:21&$3520.0$&$94.6$&$16.1$&$659.6\pm11.5$&$137.9\pm31.5$&$16.6\pm1.5$&$5.7$\\
80037-01-01-01&11-04-2003&21:30&$3520.0$&$94.0$&$16.1$&$632.4\pm13.4$&$119.1\pm35.4$&$13.9\pm1.6$&$4.3$\\
80037-01-01-02&11-05-2003&14:51&$3504.0$&$109.5$&$16.3$&$521.5\pm18.8$&$152.1\pm53.0$&$14.4\pm2.3$&$3.1$\\
80037-01-01-04&11-04-2003&23:05&$3504.0$&$119.0$&$16.9$&$641.3\pm22.5$&$146.7\pm52.3$&$14.3\pm2.0$&$3.5$\\
80037-01-05-00&02-04-2004&09:23&$3024.0$&$88.6$&$16.5$&$660.0\pm22.8$&$129.7\pm37.9$&$15.4\pm1.9$&$4.1$\\
80037-01-05-00&02-04-2004&10:57&$3392.0$&$89.2$&$16.5$&$656.9\pm12.0$&$122.6\pm34.1$&$17.4\pm1.8$&$4.9$\\
80037-01-05-00&02-04-2004&12:31&$3456.0$&$67.6$&$16.5$&$669.6\pm16.1$&$108.1\pm39.3$&$17.3\pm2.5$&$3.5$\\
80037-01-05-01&02-05-2004&09:00&$3040.0$&$72.0$&$17.0$&$703.1\pm10.5$&$69.8\pm36.5$&$14.9\pm2.5$&$3.0$\\
\hline
\hline
\end{tabular}
\end{center}
\end{table*}
\setcounter{table}{0}
\begin{table*}
\caption{Continued}
\begin{center}
\begin{tabular}{cccccccccccc}
\hline
\hline
80037-01-06-00&03-12-2004&07:01&$3344.0$&$160.9$&$30.0$&$698.2\pm2.1$&$36.2\pm5.4$&$12.9\pm0.0$&$9.4$\\
80037-01-06-00&03-12-2004&08:36&$3392.0$&$164.5$&$30.0$&$720.3\pm2.2$&$45.7\pm6.1$&$14.2\pm0.7$&$10.5$\\
80037-01-06-01&03-13-2004&09:49&$3376.0$&$124.7$&$17.2$&$810.7\pm10.7$&$123.7\pm40.3$&$14.3\pm1.6$&$4.6$\\
80037-01-06-03&03-15-2004&09:05&$3392.0$&$185.0$&$16.5$&$1179.6\pm10.9$&$52.2\pm17.0$&$7.0\pm0.9$&$3.8$\\
80037-01-06-03&03-15-2004&10:40&$3392.0$&$187.6$&$16.5$&$1188.6\pm9.8$&$73.8\pm27.1$&$8.1\pm1.0$&$3.9$\\
80037-01-07-04&03-20-2004&13:38&$2288.0$&$141.2$&$15.8$&$735.9\pm1.4$&$22.5\pm4.2$&$11.3\pm0.7$&$7.7$\\
80037-01-08-00&04-26-2004&03:08&$3392.0$&$108.5$&$16.9$&$607.5\pm12.1$&$103.9\pm27.1$&$14.0\pm1.5$&$4.7$\\
80037-01-08-00&04-26-2004&04:42&$3456.0$&$107.6$&$16.9$&$580.9\pm23.0$&$157.4\pm64.0$&$14.9\pm1.9$&$3.9$\\
80037-01-08-00&04-26-2004&06:20&$3200.0$&$109.3$&$16.9$&$599.8\pm14.3$&$102.0\pm32.4$&$12.8\pm1.6$&$4.0$\\
80037-01-08-00&04-26-2004&07:59&$2976.0$&$110.1$&$16.9$&$621.7\pm12.8$&$95.6\pm30.8$&$12.9\pm1.6$&$4.1$\\
80037-01-08-02&04-28-2004&05:33&$3376.0$&$144.0$&$16.2$&$791.8\pm11.9$&$126.8\pm42.8$&$14.5\pm1.7$&$4.3$\\
80037-01-08-02&04-28-2004&07:12&$3088.0$&$142.1$&$16.2$&$775.5\pm10.4$&$121.3\pm44.8$&$15.3\pm1.9$&$4.0$\\
80037-01-08-03&04-29-2004&06:48&$3200.0$&$111.7$&$16.4$&$562.7\pm19.2$&$111.4\pm45.9$&$13.3\pm2.2$&$3.0$\\
80037-01-08-04&04-27-2004&05:58&$3216.0$&$162.3$&$16.2$&$601.0\pm1.2$&$9.9\pm3.2$&$6.7\pm0.8$&$4.5$\\
80037-01-08-04&04-27-2004&07:35&$3024.0$&$155.8$&$16.2$&$876.1\pm23.7$&$108.5\pm47.1$&$10.1\pm1.7$&$3.0$\\
80037-01-08-06&04-25-2004&08:23&$2832.0$&$78.2$&$15.8$&$462.2\pm14.9$&$134.9\pm39.5$&$17.5\pm1.9$&$4.5$\\
80037-01-09-00&04-30-2004&04:48&$3392.0$&$122.9$&$16.0$&$698.1\pm12.1$&$111.4\pm33.9$&$12.9\pm1.4$&$4.6$\\
80037-01-09-00&04-30-2004&06:24&$3328.0$&$124.1$&$16.0$&$685.0\pm8.9$&$84.9\pm23.8$&$12.4\pm1.3$&$4.8$\\
80037-01-09-00&04-30-2004&08:02&$3088.0$&$126.3$&$16.0$&$702.0\pm9.1$&$101.5\pm35.1$&$13.5\pm1.5$&$4.4$\\
80037-01-09-01&05-04-2004&01:45&$2960.0$&$154.2$&$16.3$&$850.9\pm15.2$&$121.2\pm47.8$&$12.5\pm1.7$&$3.7$\\
80037-01-09-01&05-04-2004&04:54&$3392.0$&$115.5$&$16.3$&$595.4\pm2.7$&$18.4\pm8.3$&$8.4\pm1.3$&$3.3$\\
80037-01-09-01&05-04-2004&06:28&$3440.0$&$160.9$&$16.3$&$563.8\pm3.2$&$20.2\pm9.3$&$6.8\pm1.1$&$3.2$\\
80037-01-09-02&05-01-2004&04:28&$3280.0$&$118.6$&$16.1$&$621.4\pm11.7$&$89.4\pm37.6$&$13.1\pm2.0$&$3.3$\\
80037-01-09-02&05-01-2004&06:01&$3376.0$&$117.7$&$16.1$&$622.2\pm9.5$&$68.4\pm24.9$&$12.3\pm1.7$&$3.7$\\
80037-01-09-02&05-01-2004&07:39&$3152.0$&$120.5$&$16.1$&$650.1\pm14.1$&$75.4\pm28.3$&$11.4\pm1.7$&$3.4$\\
80037-01-09-03&05-05-2004&01:23&$2832.0$&$186.7$&$16.1$&$741.3\pm3.7$&$59.6\pm11.0$&$12.0\pm0.7$&$8.0$\\
80037-01-09-03&05-05-2004&02:57&$3040.0$&$158.2$&$16.1$&$784.3\pm13.9$&$111.1\pm38.3$&$10.7\pm1.3$&$4.1$\\
80037-01-09-04&05-03-2004&05:16&$3376.0$&$114.6$&$17.6$&$616.5\pm18.0$&$124.1\pm41.1$&$14.9\pm2.0$&$3.7$\\
80037-01-09-04&05-03-2004&06:51&$3392.0$&$116.0$&$17.6$&$628.4\pm18.3$&$128.9\pm36.4$&$15.6\pm1.8$&$4.4$\\
80037-01-10-00&05-08-2004&08:09&$3392.0$&$121.2$&$17.2$&$676.4\pm11.7$&$121.8\pm31.3$&$14.4\pm1.4$&$5.1$\\
80037-01-10-00&05-08-2004&09:46&$1712.0$&$120.7$&$17.2$&$690.3\pm29.4$&$172.1\pm80.0$&$15.0\pm2.5$&$3.0$\\
80037-01-10-01&05-07-2004&10:10&$3120.0$&$186.9$&$16.6$&$699.9\pm2.5$&$30.0\pm5.4$&$10.7\pm0.7$&$7.6$\\
80051-02-01-00&01-18-2004&12:21&$3376.0$&$162.5$&$16.1$&$744.0\pm2.5$&$29.3\pm8.2$&$9.9\pm0.9$&$5.4$\\
80051-02-01-00&01-18-2004&13:59&$3152.0$&$181.1$&$16.1$&$676.8\pm1.5$&$21.0\pm5.2$&$9.6\pm0.7$&$6.4$\\
80051-02-01-01&01-19-2004&10:24&$3440.0$&$138.6$&$16.0$&$643.2\pm4.4$&$28.6\pm13.9$&$8.2\pm1.3$&$3.1$\\
80051-02-01-01&01-19-2004&11:59&$3392.0$&$124.0$&$16.0$&$852.5\pm20.8$&$132.5\pm63.1$&$13.5\pm2.1$&$3.2$\\
80051-02-01-02&01-20-2004&11:38&$3392.0$&$190.3$&$16.0$&$709.6\pm3.6$&$41.6\pm10.3$&$9.9\pm0.8$&$5.8$\\
80051-02-01-02&01-20-2004&13:13&$3392.0$&$146.4$&$16.0$&$704.3\pm1.3$&$16.6\pm3.4$&$9.9\pm0.7$&$6.9$\\
80051-02-01-03&01-21-2004&14:30&$3168.0$&$147.8$&$17.0$&$722.4\pm3.2$&$52.2\pm8.2$&$13.0\pm0.7$&$8.8$\\
80051-02-01-05&01-22-2004&07:50&$2704.0$&$151.9$&$16.4$&$745.6\pm1.2$&$16.1\pm3.2$&$9.3\pm0.6$&$7.3$\\
80051-02-01-05&01-22-2004&09:20&$3152.0$&$111.8$&$16.4$&$725.2\pm3.1$&$33.7\pm7.5$&$11.6\pm0.9$&$6.1$\\
80051-02-01-06&01-22-2004&12:29&$3440.0$&$168.8$&$16.6$&$1142.8\pm8.6$&$65.3\pm22.5$&$9.8\pm1.2$&$4.1$\\
80051-02-01-06&01-22-2004&14:06&$2384.0$&$158.8$&$16.6$&$767.8\pm6.9$&$49.3\pm19.8$&$10.2\pm1.4$&$3.6$\\
80051-02-02-00&01-24-2004&10:12&$3040.0$&$84.5$&$16.3$&$694.6\pm7.4$&$72.7\pm18.2$&$14.6\pm1.3$&$5.5$\\
80051-02-02-01&01-26-2004&09:28&$2912.0$&$102.6$&$30.0$&$804.5\pm10.0$&$73.3\pm28.0$&$14.7\pm1.9$&$3.8$\\
80051-02-02-01&01-26-2004&11:03&$3328.0$&$103.5$&$30.0$&$783.3\pm10.1$&$84.2\pm33.9$&$15.0\pm2.0$&$3.7$\\
80414-01-04-00&08-29-2003&20:44&$3280.0$&$142.5$&$16.7$&$522.2\pm23.7$&$218.9\pm59.4$&$17.2\pm2.1$&$4.0$\\
80414-01-04-00&08-29-2003&22:18&$3456.0$&$143.7$&$16.7$&$557.7\pm18.5$&$172.6\pm54.9$&$15.5\pm1.6$&$4.7$\\
80414-01-04-01&08-30-2003&18:49&$2800.0$&$201.3$&$18.0$&$767.6\pm10.6$&$99.9\pm28.5$&$14.0\pm1.4$&$4.9$\\
80414-01-04-01&08-30-2003&20:23&$3216.0$&$208.8$&$18.0$&$818.7\pm24.1$&$154.2\pm53.2$&$13.0\pm1.7$&$3.9$\\
80414-01-05-02&10-06-2003&23:49&$2736.0$&$110.1$&$16.9$&$761.9\pm24.2$&$193.6\pm60.1$&$16.5\pm1.9$&$4.2$\\
80414-01-06-00&10-14-2003&22:31&$3216.0$&$97.1$&$16.6$&$662.1\pm14.7$&$85.6\pm28.0$&$14.6\pm1.9$&$3.8$\\
90422-01-01-01&10-04-2004&18:23&$3264.0$&$112.9$&$15.6$&$613.2\pm17.4$&$100.6\pm34.6$&$11.4\pm1.6$&$3.5$\\
90422-01-01-01&10-04-2004&19:57&$3392.0$&$113.0$&$15.6$&$591.3\pm16.0$&$110.2\pm32.5$&$12.5\pm1.5$&$4.2$\\
90422-01-01-02&10-06-2004&19:11&$3456.0$&$122.7$&$17.6$&$640.9\pm29.3$&$216.9\pm89.0$&$15.7\pm2.2$&$3.6$\\
90422-01-01-03&10-05-2004&19:34&$3456.0$&$177.6$&$16.4$&$662.0\pm3.9$&$45.1\pm9.7$&$10.8\pm0.9$&$6.3$\\
90422-01-01-03&10-05-2004&21:09&$1056.0$&$182.4$&$16.4$&$661.2\pm1.5$&$12.8\pm3.0$&$9.5\pm0.9$&$5.5$\\
91425-01-04-03&11-17-2005&22:13&$2240.0$&$111.4$&$17.5$&$742.8\pm35.2$&$266.9\pm141.1$&$24.5\pm3.9$&$3.1$\\
\hline
\hline
\end{tabular}
\end{center}
\end{table*}
\setcounter{table}{0}
\begin{table*}
\caption{Continued}
\begin{center}
\begin{tabular}{cccccccccccc}
\hline
\hline
92411-01-01-02&09-17-2006&22:11&$3344.0$&$116.4$&$17.9$&$704.9\pm14.1$&$86.0\pm33.2$&$12.5\pm1.8$&$3.5$\\
92411-01-04-00&09-15-2006&18:18&$2960.0$&$104.2$&$18.1$&$563.6\pm43.2$&$250.3\pm120.2$&$21.1\pm3.5$&$3.1$\\
92411-01-04-01&09-16-2006&21:01&$3376.0$&$112.5$&$18.8$&$669.0\pm19.2$&$135.8\pm56.5$&$14.9\pm2.1$&$3.5$\\
92411-01-04-01&09-16-2006&22:35&$3392.0$&$115.7$&$18.8$&$716.0\pm12.0$&$77.1\pm32.1$&$12.2\pm1.8$&$3.4$\\
92411-01-06-00&10-30-2006&21:19&$2512.0$&$131.1$&$18.5$&$746.6\pm19.2$&$111.2\pm48.8$&$12.8\pm2.1$&$3.0$\\
92411-01-06-02&11-01-2006&17:10&$3376.0$&$105.8$&$15.7$&$633.7\pm33.1$&$314.2\pm127.4$&$25.7\pm3.2$&$3.9$\\
92411-01-06-03&11-02-2006&15:10&$3200.0$&$114.1$&$16.9$&$694.2\pm19.7$&$121.2\pm49.7$&$13.7\pm2.2$&$3.2$\\
92411-01-06-03&11-02-2006&16:44&$3328.0$&$119.4$&$16.9$&$733.9\pm22.0$&$268.3\pm109.1$&$20.9\pm2.4$&$4.3$\\
92411-01-06-04&10-29-2006&16:52&$3392.0$&$121.4$&$16.9$&$668.1\pm13.9$&$119.4\pm36.8$&$14.9\pm1.6$&$4.6$\\
92411-01-06-07&10-30-2006&16:35&$2192.0$&$120.5$&$16.6$&$707.3\pm14.7$&$114.2\pm45.4$&$15.7\pm2.4$&$3.3$\\
92411-01-07-00&11-05-2006&18:36&$3392.0$&$134.0$&$17.4$&$811.8\pm13.3$&$120.6\pm52.4$&$13.8\pm1.9$&$3.6$\\
92411-01-07-00&11-05-2006&20:12&$3328.0$&$137.6$&$17.4$&$853.0\pm7.8$&$58.3\pm24.9$&$10.6\pm1.6$&$3.3$\\
92411-01-07-00&11-05-2006&21:50&$3104.0$&$95.6$&$17.4$&$852.5\pm23.3$&$140.7\pm57.4$&$16.0\pm2.4$&$3.4$\\
92411-01-07-02&11-03-2006&14:44&$3072.0$&$130.3$&$16.9$&$785.4\pm9.2$&$64.7\pm30.2$&$11.5\pm1.7$&$3.4$\\
92411-01-07-02&11-03-2006&16:19&$3216.0$&$128.6$&$16.9$&$774.0\pm9.3$&$72.4\pm24.2$&$12.6\pm1.5$&$4.1$\\
92411-01-07-02&11-03-2006&17:53&$3408.0$&$125.3$&$16.9$&$759.5\pm10.1$&$66.8\pm27.1$&$11.4\pm1.6$&$3.5$\\
92411-01-07-03&11-09-2006&18:28&$3392.0$&$99.9$&$17.8$&$589.7\pm25.5$&$157.3\pm55.8$&$16.3\pm2.2$&$3.8$\\
92411-01-07-05&11-06-2006&21:23&$3136.0$&$79.2$&$16.4$&$834.5\pm17.7$&$117.9\pm44.2$&$14.3\pm1.9$&$3.7$\\
92411-01-08-00&11-10-2006&18:19&$2384.0$&$124.4$&$17.2$&$826.5\pm18.9$&$148.3\pm50.5$&$16.3\pm2.0$&$4.2$\\
92411-01-08-03&11-13-2006&13:37&$2704.0$&$126.3$&$17.9$&$798.6\pm16.5$&$110.0\pm53.9$&$13.3\pm2.2$&$3.0$\\
92411-01-08-03&11-13-2006&15:11&$3152.0$&$121.5$&$17.9$&$763.3\pm15.9$&$90.0\pm42.6$&$12.5\pm1.9$&$3.2$\\
92411-01-08-03&11-13-2006&16:45&$3392.0$&$121.2$&$17.9$&$776.8\pm20.0$&$123.2\pm48.2$&$13.1\pm1.9$&$3.5$\\
92411-01-08-03&11-13-2006&18:20&$3328.0$&$123.8$&$17.9$&$791.2\pm15.9$&$162.6\pm40.7$&$17.9\pm1.7$&$5.3$\\
92411-01-08-06&11-15-2006&19:20&$2304.0$&$116.3$&$18.2$&$693.8\pm37.8$&$221.7\pm108.9$&$20.9\pm3.3$&$3.1$\\
92411-01-12-00&02-24-2007&07:06&$2832.0$&$105.3$&$18.5$&$517.5\pm25.6$&$140.8\pm52.0$&$15.6\pm2.2$&$3.6$\\
92411-01-12-01&02-25-2007&09:49&$3360.0$&$119.5$&$18.4$&$656.1\pm12.2$&$76.8\pm33.7$&$14.4\pm2.2$&$3.3$\\
92411-01-14-01&04-10-2007&04:58&$2960.0$&$152.1$&$17.6$&$746.9\pm16.9$&$137.0\pm61.1$&$13.0\pm1.8$&$3.5$\\
93404-01-02-00&07-28-2007&00:00&$3392.0$&$121.7$&$19.0$&$702.1\pm21.9$&$115.0\pm50.5$&$15.1\pm2.4$&$3.2$\\
93404-01-04-02&09-05-2007&19:08&$2960.0$&$103.6$&$17.5$&$682.5\pm37.6$&$146.9\pm58.7$&$14.6\pm2.4$&$3.1$\\
93404-01-04-02&09-05-2007&20:43&$3328.0$&$72.3$&$17.5$&$616.6\pm16.3$&$83.5\pm36.6$&$16.5\pm2.7$&$3.0$\\
93404-01-05-01&09-13-2007&18:47&$3040.0$&$220.9$&$19.8$&$879.0\pm16.3$&$120.4\pm48.9$&$12.4\pm1.7$&$3.6$\\
93404-01-05-01&09-13-2007&20:21&$3408.0$&$229.8$&$19.8$&$854.8\pm12.2$&$75.1\pm30.2$&$9.6\pm1.4$&$3.5$\\
93404-01-05-01&09-13-2007&21:55&$3392.0$&$204.6$&$19.8$&$824.4\pm14.9$&$133.4\pm48.3$&$13.7\pm1.7$&$4.1$\\
93404-01-06-00&09-15-2007&19:28&$3264.0$&$162.9$&$16.8$&$496.1\pm17.0$&$118.1\pm47.6$&$13.8\pm1.9$&$3.6$\\
93404-01-07-00&09-22-2007&00:41&$3456.0$&$62.7$&$17.6$&$694.3\pm31.4$&$161.8\pm73.7$&$19.0\pm3.1$&$3.1$\\
93404-01-07-00&09-22-2007&02:16&$3456.0$&$120.1$&$17.6$&$681.4\pm9.3$&$100.9\pm21.9$&$14.4\pm1.2$&$6.0$\\
93404-01-09-00&10-13-2007&18:15&$3392.0$&$105.3$&$15.9$&$1221.3\pm5.4$&$28.4\pm13.4$&$7.0\pm1.2$&$3.0$\\
\hline
\hline
\end{tabular}
\end{center}
\label{boutelier_tab1}
\end{table*}

\newpage
\bigskip

\begin{table*}
\caption{The parameters of the twin QPOs together with information concerning the observations: the ObsID name (ObsID), the date of the observation (Date), the start time of the observation (Time, UTC), the observation duration (T$_{\rm obs}$), the source count rate divided by the number of active PCU units (Rate), the background count rate (also divided by the number of active PCU units) (Bkg), the frequency $\nu_{~~l,u}$, Full Width Half Maximum (FWHM$_{~l,u}$), fractional RMS amplitude (RMS$_{~l,u}$, \%) and the $R_{~l,u}$ ratio, which is computed as the Lorentzian integrated power divided by its $1\sigma$ error, for the lower and upper QPO respectively. }
\begin{center}
\tiny
\begin{tabular}{ccccccccccccccc}
\hline
\hline ObsID & Date & Time & T$_{\rm obs}$ & Rate & Bkg & $\nu_l$ & FWHM$_l$ & RMS$_l$ & R$_l$ &$\nu_u$ & FWHM$_u$ & RMS$_u$ & R$_u$   \\
\hline
\hline
10095-01-02-00&04-24-1996&14:57&$2080.0$&$163.2$&$20.9$&$560.1\pm2.1$&$13.3\pm6.8$&$6.2\pm1.0$&$3.1$&$889.4\pm10.5$&$86.9\pm34.2$&$11.0\pm1.4$&$3.8$\\
10095-01-02-00&04-24-1996&16:35&$1952.0$&$166.6$&$20.9$&$603.1\pm6.2$&$42.1\pm12.7$&$9.0\pm1.1$&$4.3$&$932.4\pm18.0$&$84.0\pm0.3$&$9.4\pm1.4$&$3.4$\\
10095-01-02-00&04-25-1996&04:58&$3456.0$&$163.5$&$20.9$&$584.2\pm4.0$&$29.2\pm10.8$&$7.2\pm0.9$&$3.8$&$910.4\pm8.8$&$80.3\pm22.3$&$10.3\pm1.0$&$4.9$\\
10095-01-03-000&08-06-1996&22:29&$3456.0$&$159.5$&$18.8$&$680.5\pm1.6$&$26.4\pm4.6$&$10.3\pm0.6$&$8.2$&$993.7\pm14.0$&$85.3\pm37.3$&$8.7\pm1.3$&$3.3$\\
10095-01-03-000&08-07-1996&00:09&$3216.0$&$151.7$&$18.8$&$628.2\pm6.7$&$54.9\pm17.5$&$9.4\pm1.1$&$4.4$&$944.4\pm13.3$&$108.0\pm36.3$&$11.3\pm1.3$&$4.4$\\
10095-01-03-000&08-07-1996&01:49&$2976.0$&$146.5$&$18.8$&$582.9\pm7.5$&$36.8\pm16.4$&$7.1\pm1.2$&$3.0$&$904.3\pm12.2$&$77.7\pm26.4$&$9.9\pm1.2$&$4.0$\\
30053-01-02-00&10-16-1998&07:28&$3520.0$&$133.3$&$17.9$&$709.8\pm4.5$&$60.9\pm17.2$&$11.8\pm1.0$&$5.8$&$1017.0\pm12.7$&$107.2\pm44.6$&$10.1\pm1.4$&$3.7$\\
30053-01-02-00&10-16-1998&09:04&$3520.0$&$129.1$&$17.9$&$648.9\pm2.0$&$27.1\pm6.4$&$9.3\pm0.7$&$6.3$&$977.4\pm13.0$&$118.4\pm47.5$&$11.1\pm1.4$&$4.0$\\
30053-01-02-00&10-16-1998&10:40&$3168.0$&$110.8$&$17.9$&$580.5\pm5.8$&$36.7\pm13.1$&$8.4\pm1.1$&$3.7$&$893.3\pm13.8$&$126.4\pm37.5$&$13.2\pm1.4$&$4.7$\\
30053-01-02-02&10-16-1998&15:28&$2880.0$&$117.3$&$16.8$&$572.3\pm4.1$&$34.4\pm12.4$&$8.8\pm1.1$&$4.1$&$896.6\pm11.6$&$72.1\pm30.6$&$9.2\pm1.4$&$3.3$\\
30053-01-02-04&10-16-1998&01:20&$2576.0$&$126.2$&$19.3$&$640.0\pm5.3$&$43.2\pm10.7$&$9.9\pm1.0$&$5.2$&$954.8\pm12.6$&$59.5\pm22.9$&$8.8\pm1.3$&$3.5$\\
30056-01-01-02&03-07-1998&10:01&$3264.0$&$144.3$&$17.4$&$741.5\pm8.0$&$84.9\pm16.6$&$11.3\pm0.9$&$6.5$&$1068.5\pm11.8$&$83.5\pm30.4$&$8.8\pm1.2$&$3.8$\\
30056-01-01-03&03-08-1998&10:20&$2144.0$&$156.2$&$16.9$&$842.9\pm19.8$&$170.1\pm73.0$&$11.8\pm1.7$&$3.5$&$1144.4\pm6.8$&$67.9\pm20.3$&$10.2\pm1.1$&$4.6$\\
30056-01-03-02&11-01-1998&15:25&$3168.0$&$152.1$&$19.2$&$686.9\pm3.6$&$45.6\pm8.4$&$10.6\pm0.8$&$6.8$&$1021.6\pm32.0$&$311.1\pm143.6$&$13.7\pm2.0$&$3.5$\\
40030-01-07-00&12-29-1998&22:02&$1520.0$&$146.2$&$17.8$&$641.5\pm1.9$&$17.5\pm5.3$&$8.1\pm0.9$&$4.7$&$951.8\pm19.4$&$115.8\pm48.5$&$10.6\pm1.6$&$3.3$\\
40429-01-03-06&10-21-1999&18:55&$3472.0$&$136.1$&$19.3$&$800.4\pm17.6$&$109.6\pm48.8$&$10.2\pm1.6$&$3.2$&$1144.0\pm9.4$&$69.4\pm25.4$&$10.7\pm1.3$&$4.1$\\
80037-01-06-00&03-12-2004&05:28&$3104.0$&$155.1$&$30.0$&$658.9\pm2.9$&$27.5\pm6.4$&$10.9\pm1.0$&$5.7$&$958.9\pm11.4$&$61.7\pm25.5$&$10.0\pm1.5$&$3.2$\\
80037-01-06-03&03-15-2004&07:30&$3248.0$&$173.3$&$16.5$&$782.7\pm12.8$&$123.5\pm41.4$&$11.1\pm1.2$&$4.5$&$1130.6\pm7.3$&$45.8\pm21.4$&$7.3\pm1.1$&$3.2$\\
80037-01-09-01&05-04-2004&03:19&$3152.0$&$162.4$&$16.3$&$591.5\pm5.6$&$35.8\pm14.7$&$8.2\pm1.2$&$3.5$&$921.3\pm12.2$&$75.4\pm25.5$&$10.0\pm1.3$&$3.9$\\
80051-02-02-00&01-24-2004&05:28&$2592.0$&$136.7$&$16.3$&$696.7\pm2.5$&$33.9\pm5.3$&$11.0\pm0.7$&$8.2$&$1008.2\pm12.9$&$71.9\pm29.0$&$8.5\pm1.3$&$3.4$\\
80051-02-02-00&01-24-2004&07:02&$2800.0$&$82.0$&$16.3$&$673.4\pm4.0$&$30.1\pm13.8$&$10.3\pm1.5$&$3.4$&$992.2\pm19.7$&$75.5\pm30.8$&$10.8\pm1.8$&$3.0$\\
80051-02-02-00&01-24-2004&08:37&$2912.0$&$78.7$&$16.3$&$638.8\pm8.7$&$56.2\pm20.6$&$12.2\pm1.6$&$3.8$&$954.4\pm12.7$&$76.3\pm34.9$&$12.2\pm1.9$&$3.2$\\
90422-01-01-03&10-05-2004&18:07&$2816.0$&$165.9$&$16.4$&$611.3\pm9.3$&$51.5\pm14.3$&$9.5\pm1.1$&$4.2$&$915.2\pm22.9$&$130.1\pm44.7$&$11.4\pm1.5$&$3.7$\\
92411-01-06-01&10-31-2006&19:13&$3200.0$&$165.1$&$18.7$&$608.7\pm3.5$&$34.4\pm11.9$&$9.7\pm1.1$&$4.5$&$928.8\pm8.2$&$69.7\pm24.1$&$10.8\pm1.3$&$4.2$\\
\hline
\hline
\end{tabular}
\end{center}
\label{boutelier_tab2}
\end{table*}

\end{document}